\documentclass[final,5p,times,twocolumn]{elsarticle}
\usepackage{graphicx}
\usepackage{amssymb,amsmath}
\usepackage[colorlinks=true]{hyperref}

\newcommand{\rmd}{\mathrm{d}}

\newcommand{\etal}{{et al. }}

\journal{Nucl. Instrum. Methods Phys. Res. B}
\begin{document}
\begin{frontmatter}

\title{Alternative scattering power for Gaussian beam model of heavy charged particles}

\author{Nobuyuki Kanematsu}
\ead{nkanemat@nirs.go.jp}
\address{Department of Accelerator and Medical Physics, 
National Institute of Radiological Sciences, 4-9-1 Anagawa, Inage-ku, Chiba 263-8555, Japan}

\begin{abstract}
This study provides an accurate, efficient, and simple multiple scattering formulation for heavy charged particles such as protons and heavier ions with a new form of scattering power that is a key quantity for beam transport in matter.
The Highland formula for multiple scattering angle was modified to a scattering-power formula to be used within the Fermi-Eyges theory in the presence of heterogeneity.
An analytical formula for RMS end-point displacement in homogeneous matter was also derived for arbitrary ions. 
The formulation was examined in terms of RMS angles and displacements in comparison with other formulations and measurements.
The results for protons, helium ions, and carbon ions in water agreed with them at a level of 2\% or the differences were discussed.
\end{abstract}


\end{frontmatter}

\section{Introduction}

Theory of Coulomb scattering was developed in detail many years ago.
Rutherford \cite{Rutherford1911} studied elastic scattering of a charged particle by a point-like nucleus and found $1/\theta^4$ behavior in the distribution of scattering angle $\theta$.
In matter, a particle undergoes multiple processes of scattering by nuclei with electric field screened by orbital electrons. 
Moli\`{e}re developed a rigorous theory for such a system and formulated an analytical expression of the angular distribution for a particle interacting with target atoms \cite{Moliere1948}.
The resultant Moli\`{e}re distribution has Gaussian behavior at small angles and $1/\theta^4$ behavior at large angles. 
The Gaussian behavior is formed by multiple small-angle scatters with the central-limit theorem in statistics, while the $1/\theta^4$ tail reflects single large-angle scattering.

In cases of accelerated particles, either whole or subset of them are often modeled as a Gaussian beam because it is physically approximate to the reality, algorithmically efficient for good localization nature, and numerically easy with the standard math library.
Fermi and Eyges \cite{Eyges1948} developed a theory for Gaussian beam transport in matter, where number density $\Phi$ at transverse position $y$ and angle $\theta$ for a $N$-particle system is described as
\begin{equation}
\Phi(y,\theta) = \frac{N}{2 \pi} \sqrt{\frac{1}{\overline{y^2}\,\overline{\theta^2}-\overline{y\theta}^2}} \exp\!\left(-\frac{1}{2}\,\frac{\overline{\theta^2}\,y^2-2\,\overline{y\theta}\,y\theta+\overline{y^2}\,\theta^2}{\overline{y^2}\,\overline{\theta^2}-\overline{y\theta}^2}\right),
\label{eq:phase-space}
\end{equation}
which is characterized by angular variance $\overline{\theta^2} = \sum_{i=1}^N \theta_i^2/N$, spatial variance $\overline{y^2} = \sum_{i=1}^N y_i^2/N$, and covariance $\overline{y\theta} = \sum_{i=1}^N y_i \theta_i/N$.
As the particles undergo energy loss and multiple scattering, they are modified as
\begin{eqnarray}
\overline{\theta^2}(x) &=& \int_0^x T(x')\, \rmd x', 
\label{eq:int0}
\\
\overline{y \theta}(x) &=& \int_0^x (x-x')\, T(x')\, \rmd x', 
\label{eq:int1}
\\
\overline{y^2}(x) &=& \int_0^x (x-x')^2\, T(x')\, \rmd x',
\label{eq:int2}
\end{eqnarray}
where scattering power $T$ is the key quantity that drives the beam development in heterogeneous system along longitudinal position $x$. 
Note that we consistently deal with projected position $y$ and angle $\theta$ in this work and that $\overline{\theta^2}$, $\overline{y\theta}$, $\overline{y^2}$, $T$, and constant $E_\mathrm{s}^2$ (Sec.~\ref{sec:theories}) are thus 1/2 of the conventional definitions with radial positions and polar angles.
We also give specific symbols to such quantities later in various formulations.

For a given model of $\overline{\theta^2}(x)$, it is possible to numerically obtain effective scattering power $\tilde{T} = [\overline{\theta^2}(x+\Delta x)-\overline{\theta^2}(x)]/\Delta x$ for small step $\Delta x$ \cite{Russell1995,Kanematsu1998}.
However, it is desirable to have an analytical formula for the scattering power, which is the main objective of this work, not only for theoretical cleanness but also for further analytical derivation of physical quantities.

In the following sections, we examine various Gaussian approximation models for multiple scattering, propose a new form of scattering power as a better solution, and compare with other studies in terms of resultant physical quantities such as angles and displacements.
Although this work is primarily intended for heavy charged particle radiotherapy with protons or heavier ions, the subject is substantially general and may be useful for other beam-transport applications \cite{Kaneda2007,Neri2004}. 

\section{Materials and methods}

\subsection{Gaussian approximations for multiple scattering}\label{sec:theories}

\paragraph{Fermi-Rossi (FR) formulation}

Fermi and Rossi \cite{Rossi1941} developed a theory of multiple scattering with RMS angle $\theta_{\rm FR}$ and scattering power $T_{\rm FR}$ formulated as
\begin{equation}
T_{\rm FR} = \frac{\rmd \theta_{\rm FR}^2}{\rmd x} = \frac{E_\mathrm{s}^2}{X_0} \left( \frac{ z}{p v}\right)^2,
\label{eq:rossi}
\end{equation}
where $E_\mathrm{s} = m_e c^2 \sqrt{2\pi/\alpha} \approx 15.0$ MeV is a constant energy, $z$, $p$, and $v$ are the charge$/e$, the momentum, and the speed of the particle, and $X_0$ is the radiation length that conveniently encapsulates the material properties \cite{Yao2006}.
Since the F-R formulation relies on the central-limit theorem, existence of the $1/\theta^4$ tail at large angles leads to inaccuracy.
Nevertheless, the FR or equivalent form of $T \propto (z/pv)^2$ has been commonly used in the Fermi-Eyges theory \cite{Eyges1948,Sandison2000,Hollmark2004}.

\paragraph{Moli\`{e}re-Hanson (MH) formulation}

In the Moli\`{e}re theory, the angle distribution has a Gaussian term with RMS angle $(\chi_\mathrm{c}/\sqrt{2}) \sqrt{B}$, where $\chi_\mathrm{c}$ and $B$ can be interpreted as the characteristic angle per scattering and the mean number of scattering per particle.
The width of the central part of the Moli\`{e}re distribution is, however, slightly narrower than that of the Gaussian term due to contributions of non-Gaussian terms.
Hanson \etal \cite{Hanson1951} found the best-approximate Gaussian RMS angle for the central part as
 \begin{equation}
\theta_{\rm MH} = \frac{\chi_\mathrm{c}}{\sqrt{2}} \sqrt{B-1.2}.
\label{eq:hanson}
\end{equation}
Although the MH angle $\theta_{\rm MH}$ is often used in Gaussian beam models \cite{Russell1995,Deasy1998,Ciangaru2005}, the complexity of the theory would discourage its direct use in demanding applications and it is difficult to formulate an analytical scattering power to handle heterogeneity.
As to the details of the Moli\`{e}re theory including Fano correction, this work strictly follows the formalism by Gottschalk \etal \cite{Gottschalk1993}.

\paragraph{Integral Highland (iH) formulation}

Highland \cite{Highland1975} introduced a simple correction term to the integral form of the FR formula with an optimized energy constant,\footnote{Highland \cite{Highland1975} quantified the constant in a different form as 17.5 MeV that would be 13.9 MeV in the standard form, whereas 14.1 MeV is commonly referred to as Highland's constant and has been used as the standard.} for better agreement with the MH angle. 
Gottschalk \etal \cite{Gottschalk1993} then generalized for thick targets with formula
\begin{equation}
\theta_{\rm iH}(x) = \left(1+\frac{1}{9} \lg\frac{x}{X_0}\right) \sqrt{\int_0^x \left( \frac{14.1\,\mathrm{MeV}\,z}{p v(x')}\right)^2 \frac{\rmd x'}{X_0}},
\label{eq:gottschalk}
\end{equation}
where $\lg = \log_{10}$ is the common logarithmic function and $x$ is the thickness of a homogeneous target.
The iH angle $\theta_{\rm iH}$ was experimentally verified to be accurate \cite{Gottschalk1993,Wong1989} and has been used in practice \cite{Hong1996,Safai2008}.

Often, a target may have composite structure of multiple elements.
In such cases, use of the quadratic additivity rule $\theta_{\rm iH}^2 = \sum_i {\theta_{\rm iH}^2}_i$ may appear natural, but is incompatible with the ill-behaved logarithmic term for thin target layers to handle heterogeneity.
Kanematsu \etal \cite{Kanematsu1998} addressed the problem by further generalization
\begin{equation}
\theta_{\rm iH}(x) = \left(1+\frac{1}{9} \lg\ell(x) \right) \sqrt{ \int_0^x \left(\frac{14.1\,{\rm MeV}\,z}{pv(x')}\right)^2 \frac{\rmd x'}{X_0(x')}},
\label{eq:kanematsu}
\end{equation}
where $x$ is interpreted as the longitudinal position in the target and $\theta_{\rm iH}$ is the RMS angle growing with $x$.
Radiative path length $\ell$ is defined as
\begin{equation}
\ell(x) = \int_0^x \frac{\rmd x'}{X_0(x')},
\label{eq:ell}
\end{equation}
for any composite target or heterogeneous system with radiation length $X_0$ varying with $x$.
However, it is also difficult to derive an exact scattering power from Eq.~(\ref{eq:kanematsu}) due to the involved integral terms.

\paragraph{{\O}ver{\aa}s-Schneider ({\O}S) formulation}

{\O}ver{\aa}s \cite{Overas1960} found a good approximate relation with $pv$,
\begin{equation}
\left(\frac{p v}{p_0 v_0}\right)^2 = \left(\frac{R}{R_0}\right)^{\kappa} ,
\label{eq:overa}
\end{equation}
where $R$ is the residual range of the particle expected in water, $\kappa = 1.08$ (Sec.~{\ref{sec:r-e}) is a constant, and $p_0$, $v_0$, and $R_0$ are the initial values on the incidence.
While $R$ and $\kappa$ were originally formulated as material-dependent,\footnote{Schneider \etal \cite{Schneider2001} found strong correlation between $\kappa$ and the radiation length with fitted function $\kappa = 1.0753 + 0.12\,\exp\frac{-0.09\,\rho\,X_0}{\rm g/cm^2}$.} we determined to take water as a reference material to measure kinetic energy $E$ or $pv$.

Schneider \etal \cite{Schneider2001} applied {\O}ver{\aa}s's relation to the FR formulation and proposed a RMS angle with correction to best reproduce Gottschalk's \cite{Gottschalk1993} experimental data,
\begin{eqnarray}
\theta_{\rm {\O}S}(R) = \frac{E_\mathrm{s}\,z}{p_0\,v_0}\, \sqrt{\frac{R_0/\rho_\mathrm{S}}{(\kappa-1)\,X_0} \left\{\left(\frac{R_0}{R}\right)^{\kappa-1}-1\right\}}
\nonumber\\*
\times \sqrt{c_0 + c_1 \left(\frac{1}{2}-\frac{R}{R_0}\right)^4},
\label{eq:schneider}
\end{eqnarray}
where $c_0$, and $c_1$ are given by
\begin{equation}
c_0 = 0.888-\frac{0.00406\, \rho\, X_0}{\rm g/cm^2},
\qquad
c_1 = \frac{0.0380\,\rho\, X_0}{\rm g/cm^2}-4.86,
\end{equation}
for the target material with density $\rho$, radiation length $X_0$, and stopping-power ratio $\rho_\mathrm{S}$ with respect to water.
In the {\O}S formulation, the scattering power is analytically given by
\begin{eqnarray}
T_{\rm {\O}S}(R) &=& 
\frac{E_\mathrm{s}^2}{X_0} \left(\frac{z}{p_0\,v_0}\right)^2 \left(\frac{R_0}{R}\right)^{\kappa} 
\left[ c_0 + c_1 \left(\frac{1}{2}-\frac{R}{R_0}\right)^4 \right. \nonumber\\*
&+& \left. \frac{4\,c_1}{\kappa-1} \left(\frac{1}{2}-\frac{R}{R_0}\right)^3 \frac{R}{R_0}\left\{1-\left(\frac{R}{R_0}\right)^{\kappa-1}\right\}\right].
\label{eq:os}
\end{eqnarray}
The {\O}S formulation does not seem to be popular despite the potential adaptability to the Fermi-Eyges theory.

\subsection{Alternative scattering power}

The essence of the Highland formulation is the presence of the correction factor to the integral form of the FR formula (\ref{eq:rossi}), which is considered as a variable-separation approximation.
It is thus natural to introduce a similar correction factor $f_{\rm dH}$ in the differential 
Highland (dH) formula or the dH scattering power
\begin{equation}
T_{\rm dH} = \frac{\rmd \theta_{\rm dH}^2}{\rmd x} = f_{\rm dH}(\ell)\, \frac{E_\mathrm{s}^2}{X_0} \left(\frac{z}{p v}\right)^2 
\label{eq:dhighland}
\end{equation}
to accommodate the single-scattering effect.
The average of the factor for the entire path should coincide with the original Highland correction squared or the ratio $\theta_{\rm dH}^2/\theta_{\rm FR}^2$,
\begin{equation}
\frac{1}{\ell} \int_0^\ell f_{\rm dH}(\ell')\, \rmd \ell' = \left(1+\frac{\lg\ell}{9}\right)^2 \left(\frac{14.1\,{\rm MeV}}{E_\mathrm{s}}\right)^2,
\label{eq:corrint}
\end{equation}
leading to a simple solution,
\begin{eqnarray}
f_{\rm dH}(\ell) &=& \left(\frac{14.1\,{\rm MeV}}{E_\mathrm{s}}\right)^2 \frac{\rmd}{\rmd \ell} \left[ \left(1+\frac{\lg \ell}{9}\right)^2 \ell \right]\nonumber\\*
&=& \left(\frac{14.1\,{\rm MeV}}{E_\mathrm{s}}\right)^2 \left(1+\frac{\lg\ell}{9}\right)
\left( 1+\frac{2}{9 \ln 10}+\frac{\lg\ell}{9}\right) \nonumber\\*
&\approx& 0.970 \left(1+\frac{\ln\ell}{20.7}\right) \left(1+\frac{\ln\ell}{22.7}\right),
\label{eq:corr}
\end{eqnarray}
with which $T_{\rm dH}$ is exactly defined.

\begin{figure}
\includegraphics{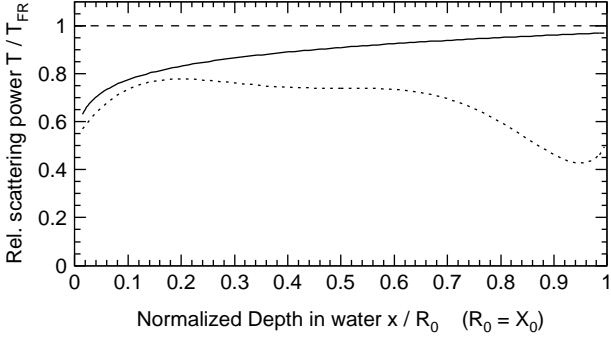}
\caption{The differential Highland (solid) and the {\O}ver{\aa}s-Schneider (dotted) scattering powers relative to the Fermi-Rossi scattering power (dashed) of water as a function of normalized depth for particles with incident range $R_0 = X_0$ .}
\label{fig:scat}
\end{figure}

Figure \ref{fig:scat} shows relative strength of scattering powers $T_{\rm FR}$, $T_{\rm dH}$, and $T_{\rm {\O}S}$ of water ($\rho = \rho_\mathrm{S} = 1$, $X_0 = 36.08$ cm) as a function of normalized depth $\rho_\mathrm{S} x/R_0 = 1-R/R_0$ for particles with incident range $R_0 = X_0$.
These curves exhibit significant differences reflecting the complexity of the formulations.

\subsection{Beam development}\label{sec:development}

\subsubsection{Range--energy relation}\label{sec:r-e}

For a semi-relativistic heavy charged particle, the Bethe theory describes its mean stopping behavior in matter \cite{ICRU1993}.
 At fixed velocity $v$, particle dependency of stopping power $S = -\rmd E/\rmd x \propto z^2$ and that of kinetic energy $E \propto $ mass $m$ lead that of residual range to $R = \int_0^{E} \rmd E'/S(E') \propto m/z^2$.
Relation $pv = E(E+2mc^2)/(E+mc^2)$ leads to $pv \propto m$.
These extend {\O}ver{\aa}s's relation for fully stripped ions or nuclei with charge $z\,e$ and mass $m = A\,u$ as
 \begin{equation}
\left(\frac{m_p}{m}\frac{pv}{\rm MeV}\right)^2 = \left(\frac{m_p}{m}\frac{z^2\, R}{\lambda \,{\rm cm}}\right)^{\kappa},
\quad 
\left(\begin{array}{c}\kappa\\ \lambda\end{array}\right) =
\left(\begin{array}{c}1.08\\ 4.67 \times 10^{-4}\end{array}\right)
\label{eq:r-pv}
\end{equation}
where $m_p = 1.0073\,u$ is the proton mass and $u = 931.5\,{\rm MeV}/c^2$ is the atomic mass unit.
Parameters $\kappa$ and $\lambda$ were determined with ICRU \cite{ICRU1993} data points for protons in water with $(E/{\rm MeV}, R/{\rm cm}) = (200, 25.96)$ and $(400, 82.25)$.

\subsubsection{Numerical computation for heterogeneous systems}

The beam development is computed in a stepwise manner to deal with particle energy loss and medium heterogeneity.
In small step $\Delta x$ from $x$, the residual range and the radiative path length are modified by
\begin{equation}
\Delta R = -\rho_\mathrm{S}(x)\, \Delta x,
\qquad \Delta \ell = \frac{\Delta x}{X_0(x)},
\end{equation}
and integrals (\ref{eq:int0})--(\ref{eq:int2}) are translated into increments
\begin{eqnarray}
\Delta\overline{\theta^2} &=& \tilde{T}\,\Delta x,
\label{eq:dqq}\\
\Delta \overline{y\theta} &=& \left(\overline{\theta^2}+\frac{\tilde{T}}{2}\,\Delta x\right)\Delta x,
\label{eq:dyq}\\
\Delta \overline{y^2} &=& \left[2\,\overline{y\theta}+\left(\overline{\theta^2}+\frac{\tilde{T}}{3}\,\Delta x\right)\Delta x \right]\Delta x.
\label{eq:dyy}
\end{eqnarray}
Effective scattering power $\tilde{T}$ in the FR formulation is 
\begin{equation}
\tilde{T}_{\rm FR} = \frac{\ E_\mathrm{s}^2}{X_0(x)}\, \frac{z^2}{pv(x)\ pv(x+\Delta x)},
\label{eq:tefr}
\end{equation}
where the geometric-mean $pv$ represent its effective value accurately for steps within 20\% of the residual range \cite{Gottschalk1993}.
The effective {\O}S and dH scattering powers are 
\begin{eqnarray}
\tilde{T}_{\rm {\O}S} &=& \left[ c_0 + c_1 \left(\frac{1}{2}-\frac{\tilde{R}}{R_0}\right)^4 \right. \nonumber\\*
&+& \left. \frac{4\,c_1}{\kappa-1} \left(\frac{1}{2}-\frac{\tilde{R}}{R_0}\right)^3 \frac{\tilde{R}}{R_0}\left\{1-\left(\frac{\tilde{R}}{R_0}\right)^{\kappa-1}\right\}\right] \tilde{T}_{\rm FR},
\label{eq:teos}
\\
\tilde{T}_{\rm dH} &=& 0.970 \left(1+\frac{\tilde{\ell}}{20.7}\right) \left(1+\frac{\tilde{\ell}}{22.7}\right)\, \tilde{T}_{\rm FR},
\label{eq:tehk}
\end{eqnarray}
where $\tilde{R} \approx R+\Delta R/2$ and $\tilde{\ell} \approx \ell+\Delta\ell/2$ are approximate effective values for the step.

Since the $R$--$E$ relation may have 1\% or more uncertainty, we limit depth step $\rho_\mathrm{S}\,\Delta x$ to be more than 0.5\% of the initial range $R_0$ to balance accuracy and efficiency.
In addition, we limit the step to be less than 10\% of the residual range $R$ and within distance $\delta$ to the next heterogeneous layer boundary, by
\begin{equation}
\Delta x = \begin{cases}
\min\left(\delta, \max\left(\frac{0.005\,R_0}{\rho_\mathrm{S}}, \frac{0.1\,R}{\rho_\mathrm{S}} \right)\right) & \text{for}\quad R > 0.01 R_0 \\
R/\rho_\mathrm{S} & \text{for}\quad R \le 0.01 R_0.
\end{cases}
\end{equation}

In the last step $\Delta x = R/\rho_\mathrm{S}$ to the end point, $\tilde{T}$ would diverge as $pv \to 0$ in Eq.~(\ref{eq:tefr}).
Since angle $\theta$ loses its physical significance there, we should only deal with $\overline{y^2}$.
The last increment $\Delta \overline{y^2}_0$ can be analytically calculated with Eqs.~(\ref{eq:int2}), (\ref{eq:dhighland}), and (\ref{eq:r-pv}), as
\begin{eqnarray}
\Delta \overline{y^2}_0 &=& \int_0^R \left(\frac{R'}{\rho_\mathrm{S}}\right)^2\,T(R') \, \frac{\rmd R'}{\rho_\mathrm{S}}
\label{eq:lastdyy} \\*
&=& \frac{f_{\rm dH}}{X_0} 
\left(\frac{E_\mathrm{s}\,z}{\rm MeV}\, \frac{m_p}{m}\right)^2 \left(\frac{R}{ \rho_\mathrm{S}}\right)^3
\frac{1}{3-\kappa} 
\left(\frac{m_p}{m}\, \frac{z^2\, R}{\lambda\, {\rm cm}}\right)^{-\kappa} \nonumber
\end{eqnarray}
using conversion $x' \to R' = R_0-\rho_\mathrm{S}\,x'$.

\subsubsection{Analytical formula for homogeneous systems}

Analytical integral (\ref{eq:lastdyy}) for the last step is valid for larger distances in the absence of heterogeneity.
Suppose an infinitesimal parallel beam with initial range $R_0$ is incident into a homogeneous target, the particles traverse distance $R_0/\rho_\mathrm{S}$ and stop with mean square displacement $\overline{y^2}_0$.
The effective Highland correction factor to $\overline{y^2}_0$ should be $R^2$-weighted mean of $f_{\rm dH}(\ell)$ in Eq.~(\ref{eq:corr}) with $\ell = (R_0-R)/(\rho_\mathrm{S} X_0)$ as
\begin{equation}
f_{y^2 0} = \frac{3}{R_0^3} \int_0^{R_0} f_{\rm dH}\, R^2 \rmd R 
\approx 0.816 \left(1+ \frac{1}{9.95} \ln\frac{R_0}{\rho_\mathrm{S} X_0}\right).
\end{equation}
We thus obtain analytical RMS end-point displacement
\begin{equation}
\sigma_{y 0}(R_0) = \frac{E_\mathrm{s}}{\rm MeV} \sqrt{\frac{f_{y^2 0}(R_0)}{(3-\kappa)\, X_0 }} \left(\frac{R_0}{\rho_\mathrm{S}}\right)^\frac{3}{2}
 \left(\frac{R_0}{\lambda\, {\rm cm}}\right)^{-\frac{\kappa}{2}} z^{1-\kappa} \left(\frac{m}{m_p}\right)^{\frac{\kappa}{2}-1} 
\label{eq:sigmay}
\end{equation}
for stopping ions in a homogeneous system.

\subsection{Application and validation}

\paragraph{Range--energy relation}

Since this work is heavily dependent on the $R$--$E$ (or $pv$) relation, we first examined its accuracy against the standard data \cite{ICRU1993} for the interested energy region of $E/A \lesssim 400$ MeV in comparison with $R$--$E$ relation $R/{\rm cm} = 0.0022\, (E/{\rm MeV})^{1.77}$ similarly proposed by Bortfeld \cite{Bortfeld1997} for $E \lesssim 250$ MeV protons.

\paragraph{Scattering angle}

\begin{table}
\caption{Atomic properties (mass density, mass-electron density, mean excitation energy, radiation mass length, and effective mass-stopping-power ratio) of water and target materials.}
\label{tab:materials}
\begin{tabular}{lrrrrr}
\hline\hline
Material & $\rho/{\rm \frac{g}{cm^3}}$ & $\frac{n_e}{\rho N_\mathrm{A}}/{\rm\frac{mol}{g}}$ & $I/{\rm eV}$ & $\rho X_0/{\rm\frac{g}{cm^2}}$ & ${\frac{\tilde{\rho}_\mathrm{S}}{\rho}}/{\rm\frac{cm^2}{g}}$ \\
\hline
Water & 1 & 0.5551 & 75 & 36.08 & 1 \\
Beryllium & 1.85 & 0.4438 & 63.7 & 65.19 & 0.8195 \\
Copper & 8.96 & 0.4564 & 322 & 12.86 & 0.6674 \\
Lead & 11.35 & 0.3958 & 823 & ~~6.37 & 0.4913 \\
\hline\hline
\end{tabular}
\end{table}

RMS angles of the various formulations were compared against the reference MH angles calculated by Gottschalk \etal \cite{Gottschalk1993} for 158.6 MeV protons incident into beryllium, copper, and lead targets with properties in Table~\ref{tab:materials}, where the effective mass-stopping-power ratio $\tilde{\rho}_\mathrm{S}/\rho$ of the materials were derived from the 158.6 MeV proton ranges in the targets \cite{Gottschalk1993} and that in water $R_0 = 17.30$ cm estimated by relation (\ref{eq:r-pv}).

\paragraph{Transverse displacement}

RMS transverse displacements $\sigma_y$ in water ($\rho_\mathrm{S} = 1$, $X_0 = 36.08$ cm) were calculated with the FR, {\O}S, and dH scattering powers as a function of depth $x$ for projectile nuclei with incident range $R_0 = 29.4$ cm $^1{\rm H}$, 29.4 cm $^4{\rm He}$, and 29.7 cm $^{12}{\rm C}$ to compare with Phillips's measurements of 1/e radius $\sqrt{2}\,\sigma_y$ \cite{Hollmark2004}.

\paragraph{End-point displacement}

Similarly, the RMS transverse displacements at the end point, $\sigma_{y 0}$, of $^1{\rm H}$, $^4{\rm He}$, and $^{12}{\rm C}$ nuclei incident into water were calculated with varied incident ranges and were compared with the analytical formula (\ref{eq:sigmay}) and measurements. 

In addition to Phillips's measurements \cite{Hollmark2004}, we included two proton points measured by Preston and Kohler in their unpublished work in 1968, which were $\sqrt{2}\,\sigma_{y 0}= (0.346\pm0.009)$ cm for $R_0 = 11.4$ cm and $\sqrt{2}\,\sigma_y = (0.368\pm0.010)$ cm at $x = 12.4$ cm for $R_0 = 12.8$ cm converted to $\sqrt{2}\,\sigma_{y 0} = 0.391$ cm with their universal curve
\begin{equation}
\frac{\sigma_y(x)}{\sigma_{y 0}(R_0)} = \sqrt{3 x_R^2 -2 x_R -2 (1-x_R)^2 \ln (1-x_R)},
\label{eq:pk}
\end{equation}
where $x_R = \rho_\mathrm{S}\,x/R_0$ is the normalized depth.
We also added other two points calculated with the Hanson form of Moli\`{e}re's theory by Deasy \cite{Deasy1998} for 160 and 250 MeV protons. 
Deasy evaluated transversal FWHM's of Bragg peak as $2\sqrt{2\ln 2}\, \sigma_y = 0.91$ cm at $x =$ 17.5 cm for $R_0 = 17.65$ cm and $2\sqrt{2\ln 2}\, \sigma_y = 1.96$ cm at unspecified depth for $R_0 = 37.94$ cm. 
The conversion factor for the former by the universal curve is 1.017 that is also applied to the latter for the best guess, leading to $\sigma_{y 0} =$ 0.393 cm and 0.846 cm.

\paragraph{Heterogeneity handling}

We calculated behaviors of $R_0 = 29.4$ cm protons in a multilayered bi-density water target, where layer $i = \{1, 2, 3, 4, ...\}$ of common thickness $t$ and density $\rho_i = \{1.1, 0.9, 1.1, 0.9, ...\}$ was placed with its upstream face at $x_i = \{0, t, 2t, 3t , ...\}$.
The RMS end-point displacements were calculated by numerical integral (\ref{eq:dyy}) using the FR, {\O}S, and dH scattering powers (\ref{eq:rossi}), (\ref{eq:os}), and (\ref{eq:dhighland}) and the effective scattering power ${{\tilde{T}}_{\rm iH}}$ per layer derived from the Gottschalk form of RMS angle $\theta_{\rm iH}$ (\ref{eq:gottschalk}),
\begin{equation}
\tilde{T}_{{\rm iH} i} = \frac{{\theta_{\rm iH}^2}_i}{t} = \left(1+\frac{1}{9}\lg\frac{\rho_i\,t}{{X_0}_\mathrm{w}}\right)^2
\int_0^t \left(\frac{14.1\,{\rm MeV}}{pv(x_i+\rmd t')}\right)^2
\frac{\rho_i\, \rmd t'}{{t\, X_0}_\mathrm{w}}.
\label{eq:layerangle}
\end{equation}
The thickness $t$ was varied in the extent of 0.01--1 cm.

\section{Results}

\paragraph{Range--energy relation}

\begin{figure}
\includegraphics{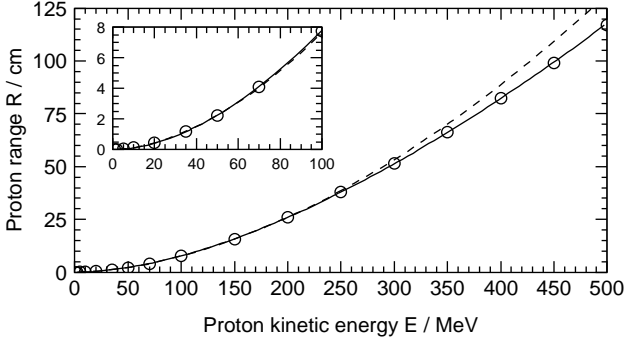}
\caption{Range--energy relation curve for protons in water translated from the $R$--$pv$ relation (solid), Bortfeld's curve (dashed), and ICRU data points ($\circ$).}
\label{fig:r-e}
\end{figure}

Figure \ref{fig:r-e} shows that the $R$--$pv$ relation (\ref{eq:r-pv}) and the standard ICRU data \cite{ICRU1993} for protons agreed within either 0.1 cm or 1\% for 0--400 MeV, which would not have been accomplished with Bortfeld's relation. 

\paragraph{Scattering angle}

\begin{figure}
\includegraphics{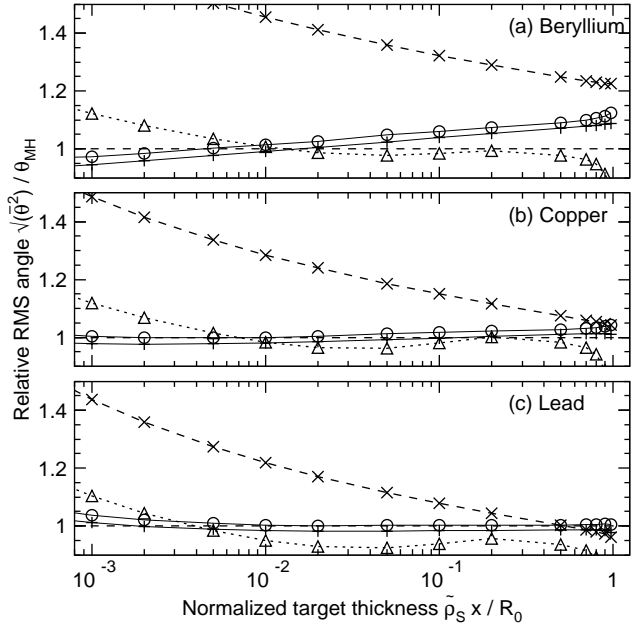}
\caption{Calculated RMS angles relative to the the Moli\`{e}re-Hanson angles $\theta_{\rm MH}$ \cite{Gottschalk1993} (dashed) by Fermi-Rossi ($\times$), integral Highland ($+$), {\O}ver{\aa}s-Schneider ($\triangle$), and differential Highland ($\circ$) formulations for 158.6 MeV protons scattered by (a) beryllium, (b) copper, and (c) lead targets as a function of normalized target thickness $\tilde{\rho}_\mathrm{S}x/R_0$.}
\label{fig:angles}
\end{figure}

\begin{table}
\caption{Calculated RMS angles for 158.6 MeV protons by beryllium, copper, and lead targets of normalized thicknesses $\tilde{\rho}_\mathrm{S}x/R_0 = 1\%$ and $10\%$, by Fermi-Rossi ($\theta_{\rm FR}$), Moli\`{e}re-Hanson ($\theta_{\rm MH}$) \cite{Gottschalk1993}, integral Highland ($\theta_{\rm iH}$), {\O}ver{\aa}s-Schneider ($\theta_{\rm {\O}S}$), and differential Highland ($\theta_{\rm dH}$) formulations.}
\label{tab:angles}
\begin{tabular}{lrrrrrr}
\hline\hline
Target & \multicolumn{2}{c}{Beryllium} & \multicolumn{2}{c}{Copper} & \multicolumn{2}{c}{Lead} \\
\cline{2-7}
$\tilde{\rho}_\mathrm{S}x/R_0$ & 1\% &10\% & 1\%& 10\% & 1\% & 10\% \\
\hline
$\theta_{\rm FR}/{\rm mrad}$ & 2.93 & 9.49 & 7.23 & 23.5 & 11.9 & 38.6 \\
$\theta_{\rm MH}/{\rm mrad}$ & 2.01 & 7.17 & 5.63 & 20.4 & 9.75 & 35.8 \\
$\theta_{\rm iH}/{\rm mrad}$ & 1.99 & 7.45 & 5.51 & 20.3 & 9.60 & 35.2 \\
$\theta_{\rm {\O}S}/{\rm mrad}$ & 2.02 & 7.06 & 5.53 & 20.0 & 9.26 & 33.6 \\
$\theta_{\rm dH}/{\rm mrad}$ & 2.04 & 7.61 & 5.63 & 20.7 & 9.78 & 35.8 \\
\hline\hline
\end{tabular}
\end{table}

The RMS scattering angles are compared in Fig.~\ref{fig:angles} and in Table \ref{tab:angles}. 
For the homogeneous target systems, the two variations of Highland angles $\theta_{\rm iH}$ and $\theta_{\rm dH}$ were in fact equally consistent with the MH angle while the FR angle deviated from them by about 50\% at small thicknesses.
The {\O}S angle was superior for the beryllium target and inferior for the copper and lead targets to the Highland angles in terms of agreement with the MH angle.

\paragraph{Transverse displacement}

\begin{figure}
\includegraphics{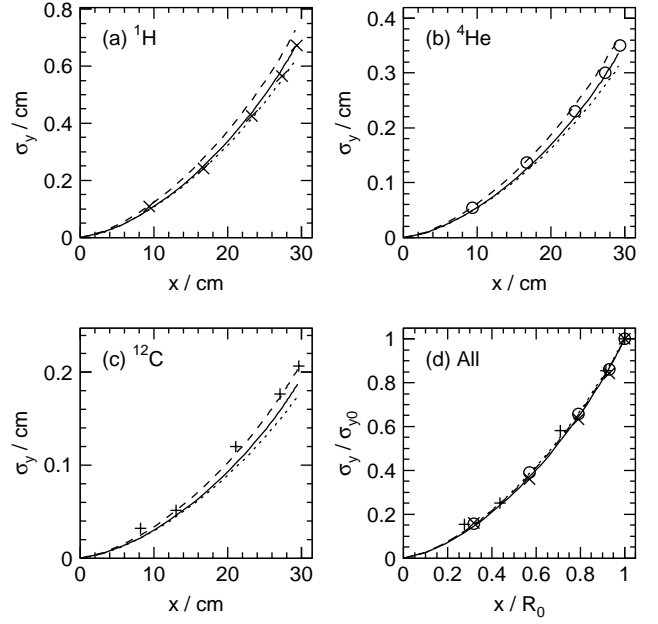}
\caption{RMS transverse displacements in water for projectiles (a) $R_0 = 29.4$ cm ${\rm ^1H}$, (b) 29.4 cm ${\rm ^4He}$, and (c) 29.7 cm ${\rm ^{12}C}$ as a function of depth, and (d) all of them in normalized scale, by numerical computation with the Fermi-Rossi (dashed), differential Highland (solid), and {\O}ve{\aa}s-Schneider (dotted) scattering powers. 
Markers indicate Phillips's measurements.}
\label{fig:displacement}
\end{figure}

Figures \ref{fig:displacement} (a)--(c) show the growths of transverse displacement of ${\rm ^1H}$, ${\rm ^4He}$, and ${\rm ^{12}C}$ nuclei in water.
In terms of relative agreement with the measurements, the dH and {\O}S formulations were excellent for ${\rm ^1H}$, the FR and dH formulations were good for ${\rm ^4He}$, and the FR was the excellent for ${\rm ^{12}C}$.
Considering their absolute scale and inherent experimental difficulties, the best would be the dH formulation with agreement within 2\% or 0.02 cm everywhere.
Figure \ref{fig:displacement} (d) shows the behaviors in the self-normalized scale, indicating that the relative displacement is very insensitive to the scattering formulations.
The universal curve (\ref{eq:pk}) would coincide with these curves.

\paragraph{End-point displacement}

\begin{figure}
\includegraphics{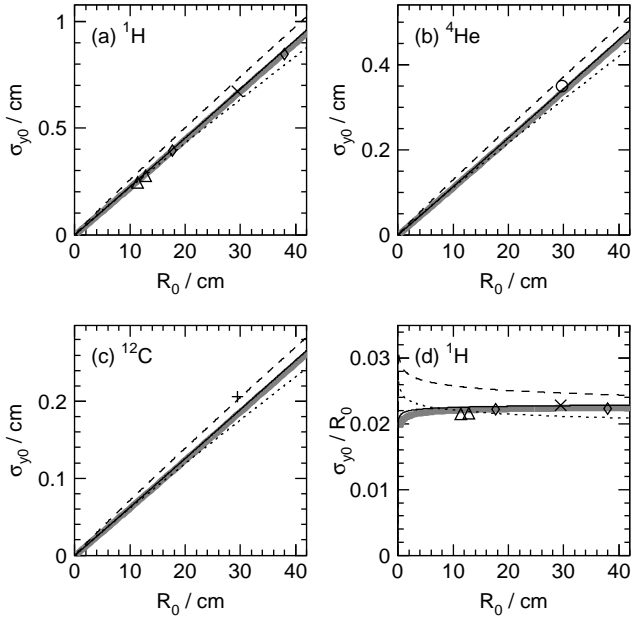}
\caption{RMS end-point displacements in water for (a) ${\rm ^1H}$, (b) ${\rm ^4He}$, and (c) ${\rm ^{12}C}$, and (d) displacement-range ratio for ${\rm ^1H}$ as a function of incident range $R_0$ by numerical computation with the Fermi-Rossi (dashed), differential Highland (solid), and {\O}ve{\aa}s-Schneider (dotted) scattering powers, the analytical formula (thick gray), measurements by Phillips ($\times$, $\circ$, $+$) and by Preston and Kohler ($\triangle$), and Moli\`{e}re-Hanson calculations by Deasy ($\diamond$).}
\label{fig:endpoint}
\end{figure}

Figures \ref{fig:endpoint} (a)--(c) show the RMS displacements of ${\rm ^1H}$, ${\rm ^4He}$, and ${\rm ^{12}C}$ nuclei at the end point in water for varied incident ranges.
In terms of agreement with the measurements, the dH formulation would be the best for the same reason as for $\sigma_y$ drawn mostly with the same data.
The analytical dH curve agreed well with the numerical computation.
As shown in Fig.~\ref{fig:endpoint} (d), behavior of the dH and {\O}S formulations were approximately linear with relation $\sigma_{y 0} \simeq 0.023\, R_0$, while the FR formulation deviated by 10\% or more with larger non-linearity.

\paragraph{Heterogeneity handling}

\begin{figure}
\includegraphics{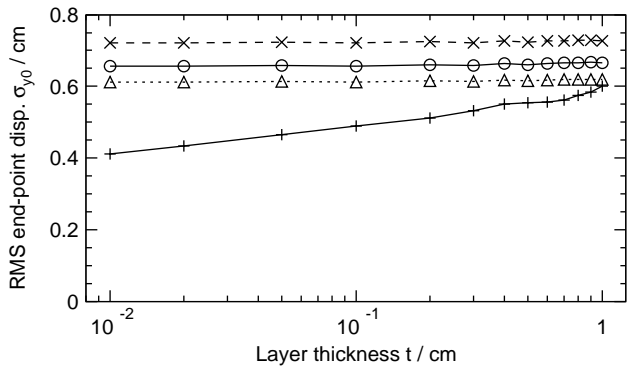}
\caption{RMS end-point displacements of $R_0 = 29.4$ cm protons in a target comprised of alternative $\rho = 1.1$ and 0.9 water layers of varied thickness $t$, calculated with the Fermi-Rossi ($\times$), {\O}ve{\aa}s-Schneider ($\triangle$), and differential Highland ($\circ$) scattering powers, and the integral Highland angle per layer ($+$).}
\label{fig:heterogeneity}
\end{figure}

Figure \ref{fig:heterogeneity} shows the behaviors of the various formulations against heterogeneity size.
The layer-wise calculation of the iH angle (\ref{eq:layerangle}) actually caused large error as addressed in Sec.~\ref{sec:theories}.
At typical layer thickness $t = 0.1$ cm, the misused iH formulation resulted in 26\% underestimation in RMS displacement with respect to the dH formulation.

\section{Discussion}

The Fermi-Eyges theory generally, smartly, and efficiently describes a Gaussian beam with a set of a few beam-defining parameters computed by path integrals.
For radiotherapy, it is useful to deal with field formation in beam delivery systems, beam customization for individual treatment targets, and beam transport in patients to give a variable pencil kernel for dose convolution algorithms. \cite{Kanematsu1998,Kanematsu2008a,Kanematsu2008b}.
The scattering power determines the accuracy of the modeled Gaussian beam. 
Fortunately, the relative error will be about a half in RMS angle and displacement because of relations $T \propto \overline{\theta^2} \propto \overline{y^2}$. 
This may be one of the reasons why the apparent differences in $T$ were not very significant in $\sigma_y$ in this work.

The FR formulation (\ref{eq:rossi}) without consideration of the single-scattering effect was inaccurate for thin targets of beryllium, copper, and lead.
The inaccuracy or the single-scattering effect decreased with the target thickness.
This is analogous to the relative decrease of Hanson's correction with increasing $B$ in Eq.~(\ref{eq:hanson}) and could be explained as follows.
Increase of the multiple scattering with thickness also increases the threshold to distinguish large-angle scattering, and thus decreases the single-scattering effect.
For particles stopping in water, the inaccuracy was in fact small.

The Highland formula is an empirical approximation of Hanson's another empirical approximation of the Moli\`{e}re theory.
There are even variations of constant parameters within the formalism \cite{Lynch1991}.
Nevertheless, the standard form adopted in this work has been experimentally validated \cite{Gottschalk1993,Wong1989,Hong1996,Szymanowski2001,Akagi2006} and thus should be a reasonable choice.
The original Highland formula, however, was designed for homogeneous systems and causes large errors when misused for fine heterogeneity.

We formulated a scattering power (\ref{eq:dhighland}) by approximate differentiation of the generalized Highland formula (\ref{eq:kanematsu}). 
The two forms were verified to be equivalent.
We analytically derived a general formula (\ref{eq:sigmay}) for RMS end-point displacement $\sigma_{y 0}$ of stopping ions in a homogeneous system, which showed surprisingly linear behavior with incident range $R_0$.
Since the formula has separate factors of $R_0$, $z$, and $m$ dependencies, one can easily compare the scattering effects for different ion beams.

The $R$--$pv$ relation in the {\O}S formulation is superior to the conventional $R$--$E$ relation and was fully utilized in this work.
As to the scattering, the {\O}S formulation generally resulted in smaller displacements in water than those of the Highland formulation by several percent, as so found by Schneider \etal \cite{Schneider2001}.
In terms of agreement with measurements by Phillips, measurements by Preston and Kohler, and MH calculations by Deasy, the {\O}S formulation was slightly inferior to the Highland formulation, although it was a fit to the other experiment.

While both formulations deal with the effect of large-angle scattering in Gaussian approximation, there is an essential difference.
Highland's correction is based on radiative path length $\ell = \int_0^x \rmd x'/X_0(x')$ that may represent the multiple scattering accumulated for the entire path.
This approach sounds reasonable because large angles can only be defined with respect to the multiple-scattering angle.
In contrast, Schneider's correction is based on relative residual range $R/R_0$ and material-specific constants $c_0$ and $c_1$, which are instantaneous quantities.
That approach seems to assume invariance of atomic composition and could be inappropriate for systems with multiple materials of different compositions.

\section{Conclusions}

We formulated a scattering power with correction for single-scattering effect based on the Highland formula.
It can be generally used within the Fermi-Eyges theory for beam applications with heavy charged particles in the presence of heterogeneity.
We derived an analytical formula for RMS displacement of ions stopped in a homogeneous system, which showed very linear behavior with the incident range in water.

The single-scattering effect was as large as 50\% for thin targets and generally decreases with thickness. 
For ions stopping in water, the effect in displacement was about 10\% and the relative growth with depth is generally insensitive to the scattering-power formulations.
The numerical and analytical calculations of this work were consistent and agreed with other experimental studies within 2\% or 0.02 cm in RMS displacement. 

\section*{Acknowledgments}

The author faithfully thanks Bernard Gottschalk for his self-published lecture notes, detailed explanations, and many insightful suggestions given to this work.
The author also thanks Uwe Schneider for a copy of his article that greatly influenced this work.


\begin{thebibliography}{00}

\bibitem{Rutherford1911}
E.~Rutherford, The scattering of $\alpha$ and $\beta$ particles by matter and the structure of the atom,
Philosophical Magazine Series 6 21 (1911) 669. 

\bibitem{Moliere1948}
G.~Moli\`{e}re, Theorie der Streuung schneller geladener Teichen II: Mehrfach- 
und Vielfachstreuung,
Z. Naturforsch. 3a (1948) 78. 

\bibitem{Eyges1948}
L.~Eyges, Multiple scattering with energy loss,
Phys. Rev. 74 (1948) 1534. 

\bibitem{Russell1995}
K.~R.~Russell, E.~Grusell, A.~Montelius, Dose calculations in proton beams: range straggling corrections and energy scaling, Phys. Med. Biol. 40 (1995) 1031. 

\bibitem{Kanematsu1998}
N.~Kanematsu, T.~Akagi, Y.~Futami, A.~Higashi, T.~Kanai, N.~Matsufuji, H.~Tomura, H.~Yamashita, A proton dose calculation code for treatment planning based on the pencil beam algorithm,
Jpn. J. Med. Phys. 18 (1998) 88. 

\bibitem{Kaneda2007}
M.~Kaneda, S.~Sato, M.~Shimizu, Z.~He, K.~Ishii, H.~Tsuchida, A.~Itoh, Energy loss and small angle scattering of swift protons passing through liquid ethanol target,
Nucl. Instr. Methods Phys. Res. B 256 (2007) 97. 

\bibitem{Neri2004}
F.~Neri, P.~L.~Walstrom, A simple empirical forward model for combined nuclear and multiple Coulomb scattering in proton radiography of thick objects,
Nucl. Instr. Methods Phys. Res. B 229 (2005) 425. 

\bibitem{Rossi1941}
B.~Rossi and K.~Greisen 1941 Cosmic ray theory, 
Rev. Mod. Phys. 13 (1941) 240. 

\bibitem{Yao2006}
W.~M.~Yao, \etal (Particle Data Group), The review of Particle Physics,
J. Phys. G 33 (2006) 1. 

\bibitem{Sandison2000}
G.~A.~Sandison, A.~V.~Chvetsov, Proton loss model for therapeutic beam dose calculations,
Med. Phys. 27 (2000) 2133. 

\bibitem{Hollmark2004}
M.~Hollmark, J.~Uhrdin, D\v{z}.~Belki\'c, I.~Gudowska, A.~Brahme, Influence of multiple scattering and energy loss straggling on the absorbed dose distributions of therapeutic light ion beams: I. Analytical pencil beam model,
Phys. Med. Biol. 49 (2004) 3247. 

\bibitem{Hanson1951}
A.~O.~Hanson, L.~H.~Lanzl, E.~M.~Lyman, M.~B.~Scott, Measurement of multiple scattering of 15.7-MeV electrons,
Phys. Rev. 84 (1951) 634. 

\bibitem{Deasy1998}
J.~O.~Deasy, A proton dose calculation algorithm for conformal therapy simulations based on Moli\`ere's theory of lateral deflections,
Med. Phys. 25 (1998) 476. 

\bibitem{Ciangaru2005}
G.~Ciangaru, J.~C.~Polf, M.~Bues, A.~R.~Smith, Benchmarking analytical calculations of proton doses in heterogeneous matter,
Med. Phys. 32 (2005) 3511. 

\bibitem{Gottschalk1993}
B.~Gottschalk, A.~M.~Koehler, R.~J.~Schneider, J.~M.~Sisterson, M.~S.~Wagner, Multiple Coulomb scattering of 160 MeV protons,
Nucl. Instr. Methods Phys. Res. B 74 (1993) 467. 

\bibitem{Highland1975}
V.~L.~Highland, Some practical remarks on multiple scattering,
Nucl. Instr. Methods 129 (1975) 497. 
V.~L.~Highland, Erratum to some practical remarks on multiple scattering,
Nucl. Instr. Methods 161 (1979) 171. 

\bibitem{Wong1989}
M.~Wong , W.~Schimmerling, M.~H.~Phillips, B.~A.~Ledewigt, D.~A.~Landis, J.~T.~Walton, S.~B.~Curtis, The multiple Coulomb scattering of very heavy charged particles,
Med. Phys. 17 (1989)163. 

\bibitem{Hong1996}
L.~Hong, M.~Goitein, M.~Bucciolini, R.~Comiskey, B.~Gottschalk, S.~Rosenthal, C.~Serago, M.~Urie, A proton beam algorithm for proton dose calculations,
Phys. Med. Biol. 41 (1996) 1305. 

\bibitem{Safai2008}
S.~Safai, T.~Bortfeld, M.~Engelsman, Comparison between the lateral penumbra of a collimated double-scattered beam and uncollimated scanning beam in proton radiotherapy,
Phys. Med. Biol. 53 (2008) 1729. 

\bibitem{Overas1960}
H.~{\O}ver{\aa}s, On small angle multiple scattering in confined bodies,
CERN Yellow Report 60-18, CERN, Geneve,1960

\bibitem{Schneider2001}
U.~Schneider, J.~Besserer, P.~Pemler, On small angle multiple Coulomb scattering of protons in the Gaussian approximation,
Z. Med. Phys. 11 (2001) 110. 

\bibitem{ICRU1993}
International Commission on Radiation Units and Measurements, Report 49 Stopping powers and ranges for protons and alpha particles, ICRU, Bethesda, MD, 1993.

\bibitem{Bortfeld1997}
T.~Bortfeld, An analytical approximation of the Bragg curve for therapeutic proton beams,
Med. Phys. 24 (1997) 2024. 

\bibitem{Kanematsu2008a}
N.~Kanematsu, S.~Yonai, A.~Ishizaki, The grid-dose-spreading algorithm for dose distribution calculation in heavy charged particle radiotherapy,
Med. Phys. 35 (2008) 602. 

\bibitem{Kanematsu2008b}
N.~Kanematsu, S.~Yonai, A.~Ishizaki, M.~Torikoshi, Computational modeling of beam-customization devices for heavy-charged-particle radiotherapy,
Phys. Med. Biol. 53 (2008) 3113. 

\bibitem{Lynch1991}
G.~R.~Lynch, O.~I.~Dahl, Approximations to multiple Coulomb scattering,
Nucl. Instr. Methods Phys. Res. B 58 (1991) 6. 

\bibitem{Szymanowski2001}
H.~Szymanowski, A.~Mazal, C.~Nauraye, S.~Biensan, R.~Ferrand, M.~Murillo, S.~Caneva, G.~Gaboriaud, J.~C.~Rosenwald, Experimental determination and verification of the parameters used in a proton pencil beam algorithm,
Med. Phys. 28 (2001) 975. 

\bibitem{Akagi2006}
T.~Akagi, N.~Kanematsu, Y.~Takatani, H.~Sakamoto, Y.~Hishikawa, M.~Abe, Scatter factors in proton therapy with a broad beam,
Phys. Med. Biol. 51 (2006) 1919. 

\end{thebibliography}
\end{document}